\documentclass{PoS}
\usepackage{cite}
\usepackage{amstext}  
\usepackage{amsfonts}
\usepackage{amsthm}
\usepackage{graphicx}
\usepackage{amsmath}
\usepackage{mathtools}
\usepackage{slashed}
\usepackage{array}                
\usepackage{xcolor} 

\newcommand{\f}[2]{\frac{#1}{#2}}

\newcommand{\ssst}[1]{\scriptscriptstyle{\text{#1}}}
\newcommand{\nosss}[1]{#1}

\newcommand{\ben}{\begin{enumerate}}
\newcommand{\een}{\end{enumerate}}
\newcommand{\bit}{\begin{itemize}}
\newcommand{\eit}{\end{itemize}}
\newcommand{\bea}{\begin{eqnarray}}
\newcommand{\eea}{\end{eqnarray}}
\newcommand{\be}{\begin{equation}}
\newcommand{\ee}{\end{equation}}
\newcommand{\ba}{\begin{align}}
\newcommand{\ea}{\end{align}}
\newcommand{\beas}{\begin{eqnarray*}}
\newcommand{\eeas}{\end{eqnarray*}}
\newcommand{\bes}{\begin{equation*}}
\newcommand{\ees}{\end{equation*}}
\newcommand{\bas}{\begin{align*}}
\newcommand{\eas}{\end{align*}}

\newcommand{\eps}{{\varepsilon}}

\newcommand{\lb}{\left(}
\newcommand{\rb}{\right)}

\newcommand{\diawidth}{1.8cm}
\newcommand{\diaheight}{1.6cm}

\newcommand{\denbar}{\bar}

\newcommand{\Dbar}[1]{D_{\nosss{#1}}}

\newcommand{\momq}{\bar{q}}
\newcommand{\tilq}{\tilde{q}}

\newcommand{\calA}{\mathcal{A}}

\newcommand{\calD}[1]{\mathcal{D}^{(#1)}(\bar q_{#1})}

\newcommand{\calN}{\mathcal{N}}

\newcommand{\calR}{\mathcal{R}}

\newcommand{\bfF}{\textbf{F}}

\newcommand{\bfK}{\textbf{K}}
\newcommand{\bfKtilde}{\tilde{\textbf{K}}}
\newcommand{\bfR}{\textbf{R}}

\newcommand{\bfKtildeloc}{\tilde{\textbf{K}}_{\mathrm{loc}}}
\newcommand{\bfS}{\textbf{S}}

\newcommand{\barN}{\bar{\mathcal{N}}}
\newcommand{\tilN}{\tilde{\mathcal{N}}}
\newcommand{\barA}{\bar{\mathcal{A}}}

\newcommand{\dendim}{D}
\newcommand{\numdim}{D_{\mathrm{n}}}

\newcommand{\singlearg}[1]{
\ifx&#1&
\else
(#1)   
\fi
}

\newcommand{\doublearg}[2]{
\ifx&#2&
(#1)  
\else
(#1,#2)   
\fi
}

\newcommand{\ampindices}[5]{{#1}_{{#2,#3}}^{#4 }\singlearg{#5}}

\newcommand{\amp}[4]{{\ampindices{\calA}{#1}{#2}{#3}{#4}}}
\newcommand{\ampbar}[4]{{\ampindices{\bar\calA}{#1}{#2}{#3}{#4}}}

\newcommand{\ratamp}[4]{{\ampindices{\delta \calR}{#1}{#2}{#3}{#4}}}
\newcommand{\deltaZ}[4]{{\ampindices{\delta Z}{#1}{#2}{#3}{#4}}}
\newcommand{\deltaZtilde}[4]{{\ampindices{\delta \tilde Z}{#1}{#2}{#3}{#4}}}

\def\tad{\mathrm{tad}}

\def\rem{\mathrm{rem}}

\newcommand{\rd}{\mathrm d}
\newcommand{\ord}{\mathcal O}

\definecolor{bluemar}{rgb}{0,0,.5}
\definecolor{redmar}{rgb}{.8,0,0}
\definecolor{greenmar}{rgb}{0,.5,0}

\renewcommand{\refeq}[1]{\mbox{\eqref{#1}}}

\newcommand{\ie}{i.e.\ }

\graphicspath{{./figures_pr/}}

\title{Rational terms in two-loop calculations}

\ShortTitle{Rational terms in two-loop calculations}

\author{Stefano Pozzorini\\
       Physik-Institut, Universit\"at Z\"urich, 
CH-8057 Z\"urich, Switzerland\\
       E-mail: \email{pozzorin@physik.uzh.ch}}

\author{Hantian Zhang\\
       Physik-Institut, Universit\"at Z\"urich, 
CH-8057 Z\"urich, Switzerland\\
       E-mail: \email{hantian.zhang@physik.uzh.ch}}

\author{\speaker{Max F. Zoller} 
\\
         Paul Scherrer Institut, Forschungsstrasse 111, 
CH-5232 Villigen PSI, Switzerland\\
        E-mail: \email{max.zoller@psi.ch}}

\abstract{
We present an extension of the renormalisation procedure based on the R-operation in $\dendim$ dimensions at two-loop level,
in which the numerators of all Feynman diagrams can be constructed in four dimensions, and the rational terms stemming from the interplay
of $(\dendim-4)$-dimensional numerator parts and UV poles are fully reconstructed from a finite set of universal local counterterms \cite{R2paper}.
This represents an extension of the concept of rational terms of type $R_2$ to two loops.
We provide a 
general method to compute one and two-loop rational counterterms from massive one-scale tadpole integrals. 
Finally, we present the full set of rational counterterms of UV origin for QED up to two-loop order.}

\FullConference{14th International Symposium on Radiative Corrections (RADCOR2019)\\ 
9-13 September 2019\\
		Palais des Papes, Avignon, France\\[10mm] Preprint PSI-PR-19-28}

\begin{document}

\section{Introduction}
Higher-order calculations of scattering amplitudes in perturbation theory are usually performed in 
$\dendim=4-2\eps$ dimensions in order to regularise divergences in Feynman integrals.
In this approach, loop momenta, $\gamma$-matrices, the metric tensor and the integration measure are defined as $\dendim$-dimensional quantities, and
divergences manifest themselves as poles in $\eps$.
This allows for a simple
Renormalisation procedure, known as the R-operation \cite{Kennedy_ROp}, which consists of the recursive subtraction of divergences stemming from
all possible sub-diagrams as well as a remaining local divergence of the full diagram. The R-operation can be implemented as the insertion of counterterms 
into diagrams of lower loop-order.

In numerical calculations, however, vectors have to be implemented in integer dimensions. 
Within automated one-loop tools such as {\sc OpenLoops} \cite{Buccioni:2019sur}, 
{\sc Recola} \cite{Denner:2017wsf},
{\sc Helac-1Loop} \cite{Bevilacqua:2011xh} or 
{\sc Madloop} \cite{Hirschi:2011pa}, the numerator of 
a Feynman integral is constructed in four dimensions, while the denominator is kept in $\dendim$ dimensions. 
The difference between one-loop numerators in
$\dendim$ and four dimensions leads to a finite contribution, which is polynomial in all external momenta
and internal masses, called a rational term of type $R_2$  
\cite{Ossola:2008xq,Draggiotis:2009yb,Garzelli:2009is,Pittau:2011qp}. 

The fully renormalised $\dendim$-dimensional amplitude is given by
\bea 
\textbf{R}\,\ampbar{1}{\gamma}{}{} &=& \ampbar{1}{\gamma}{}{}+\deltaZ{1}{\gamma}{}{} ,
\label{eq:RA1d}
\eea
where $\deltaZ{1}{\gamma}{}{}$ is the UV counterterm in the $\overline{\text{MS}}$-scheme.
Here and in the following, an $l$-loop amplitude of a Feynman diagram $\gamma$ 
constructed in $\dendim$ dimensions is denoted $\ampbar{l}{\gamma}{}{}$. An amplitude with a four-dimensional numerator, 
but $\dendim$-dimensional denominator, which is the object computed in automated tools, is denoted as $\amp{l}{\gamma}{}{}$.
The renormalised $\dendim$-dimensional amplitude can be computed from the latter, if we restore the rational contribution 
$\ratamp{1}{\gamma}{}{}$, stemming from the $(\dendim-4)$-dimensional part of the numerator,
which can be reconstructed through a counterterm insertion into a tree-level 
diagram, similar to the UV counterterm,
\bea 
\textbf{R}\,\ampbar{1}{\gamma}{}{} &=& \amp{1}{\gamma}{}{} + \deltaZ{1}{\gamma}{}{} + \ratamp{1}{\gamma}{}{} \,{}.
\label{eq:RA1f}
\eea

A two-loop amplitude $\barA_{2,\Gamma}$ corresponding to the diagram $\Gamma$ is
renormalised by first subtracting all sub-divergences and then the remaining local divergence.
This is achieved by means of one-loop counterterms 
$\deltaZ{1}{\gamma}{}{}$, associated with the divergent sub-diagrams $\gamma$, into one-loop amplitudes
$\ampbar{1}{\Gamma/\gamma}{}{}$, where $\gamma$ is contracted to a point in $\Gamma$, and a local
two-loop counterterm $\deltaZ{2}{\Gamma}{}{}$, 
\bea {\textbf{R}}\, \ampbar{2}{\Gamma}{}{} 
&=&  
\ampbar{2}{\Gamma}{}{} + \sum  \limits_{\gamma}  \deltaZ{1}{\gamma}{}{} \cdot 
\ampbar{1}{\Gamma/\gamma}{}{} + \deltaZ{2}{\Gamma}{}{} \label{eq:A2d_sep}
\,,
\eea
We generalise this procedure to
\bea
\label{eq:masterformula} 
{\textbf{R}}\, \ampbar{2}{\Gamma}{}{}   
&=&  \amp{2}{\Gamma}{}{} + 
\sum  \limits_{\gamma} \lb \deltaZ{1}{\gamma}{}{} +\deltaZtilde{1}{\gamma}{}{} + \ratamp{1}{\gamma}{}{} \rb \cdot \amp{1}{\Gamma/\gamma}{}{}
+ \lb \deltaZ{2}{\Gamma}{}{} + \ratamp{2}{\Gamma}{}{} \rb \, ,
\eea
where all Feynman integrands on the rhs
are constructed with four-dimensional numerators, and universal UV 
and rational counterterms. Here $\ratamp{1}{\gamma}{}{}$ are the well-known one-loop rational
terms\footnote{In the literature they are usually denoted as $R_2$ rational terms, 
while we will use the symbols $\ratamp{l}{}{}{}$,
where the subscript $l$ refers to the loop order.}
and
$\deltaZtilde{1}{\gamma}{}{}$ an additional renormalisation constants proportional to $\tilq^2/\eps$,
where $\tilq=\momq-q$ is the $(\dendim-4)$-dimensional part of the loop momentum.
We will show that $\ratamp{2}{\Gamma}{}{}$, which is implicitly defined in
\eqref{eq:masterformula}, can be computed from tadpole integrals with one auxiliary mass scale
and that it is indeed a rational term. Formula \eqref{eq:masterformula} can directly be implemented
in numerical algorithms and hence serves as an important step towards two-loop
automation. In this discussion we restrict ourselves to rational terms of UV origin.
A detailed description of our method and results can be found in \cite{R2paper}.
%
%
\section{Rational terms at one loop} \label{sec:oneloop}
\subsection{One loop rational terms from massive tadpoles with one scale}
Let us consider the amplitude of a one-particle irreducible
one-loop diagram $\gamma$,
\bea 
\label{eq:rtoneloopA}
\ampbar{1}{\gamma}{}{} &=& \vcenter{\hbox{\scalebox{1.}{\includegraphics[height=2.1cm]{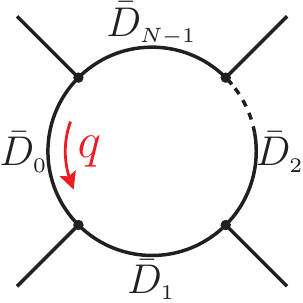}}}} \;=\;
\int\!\rd\momq_1\, \f{{\barN}(\bar{q}_1)}{\Dbar{0}(\denbar q_1)\cdots
\Dbar{N-1}(\denbar q_1)} \;
 {},
\eea
where the scalar denominators are defined as
\bea
\label{eq:rtoneloopB}
\Dbar{k}(\denbar q_1)&=& (\denbar q_1+p_k)^2-m_k^2
\eea
with external momenta $p_k$ and masses $m_k$, and the 
integration measure
\footnote{The renormalisation scale $\mu_{\overline{\ssst{MS}}}$ is related to the scale of dimensional regularisation $\mu$
via $\mu_{\overline{\ssst{MS}}}^{2\eps}=\mu^{2\eps}(4\pi)^\eps\Gamma(1+\eps)$.} 
$\rd\momq_1=(\mu_{\overline{\ssst{MS}}})^{2\eps}\rd^d\momq_1/(2\pi)^d$.

We mark $\dendim$-dimensional objects, such as loop momenta, $\gamma$-matrices and the metric, as well as their $\dendim$-dimensional Lorentz indices
with a bar, i.e.~$\momq$, $\bar\gamma^{\bar\mu}$, $\bar{g}^{\bar\mu\bar\nu}$, while their projections to four dimensions are
are denoted $q$, $\gamma^\mu$ and $g^{\mu\nu}$. The same applies to the numerator $\barN(\momq)$ of a Feynman integrand 
constructed from $\dendim$-dimensional quantities or $\calN(q)$ in four dimensions.\footnote{
In the `t~Hooft--Veltman scheme, external momenta $p_i$ 
are always defined in four dimensions.} 
The dimension of the numerator is denoted $\numdim\in\{\dendim,4\}$.

In $\numdim=\dendim$ dimensions, 
the numerator $\bar \calN(\bar q_1)$ can be split into a four-dimensional part 
\bea
\label{eq:rtoneloopE}
\calN(q_1) &=& 
\bar \calN(\bar q_1)\,\Big|_{
\bar g\to g,\, 
\bar \gamma\to \gamma\,, 
\bar q_1 \to q_1
}
\eea
and a remnant
\bea
\label{eq:rtoneloopF}
\tilde \calN(\bar q_1) &=& \barN(\bar q_1) - \calN(q_1) = \ord(\eps, \tilde q_1){}.
\eea
The rational term in \eqref{eq:RA1f} corresponds to
\bea 
\label{eq:rtoneloopI}
\ratamp{1}{\gamma}{}{} &=& \ampbar{1}{\gamma}{}{} - \amp{1}{\gamma}{}{} = 
\int\!\rd\momq_1\, \f{\tilde{\calN}(\bar q_1)}{\Dbar{0}(\denbar q_1)\cdots
\Dbar{N-1}(\denbar q_1)} \,,
\eea
which generates finite terms through the interplay of the $(\dendim-4)$-dimensional 
numerator with the $1/\eps$ UV poles.\footnote{At one loop $1/\eps$ poles of 
IR origin do not generate rational terms.
See App.~A of~\cite{Bredenstein:2008zb}.}
Hence, such $\tilN$-contributions arise only in
UV divergent 1PI diagrams.  

We now discuss a well-established method \cite{Misiak:1994zw,beta_den_comp} for the calculation of 
the UV divergent part of a Feynman diagram from massive tadpoles with one scale, and adapt it
to the calculation of rational terms.
This method is based on the exact decomposition of all propagator denominators,
\bea
\f{1}{\Dbar{i}}&=&\f{1}{\momq_1^2-M^2}+
\f{\Delta_i}{\momq_1^2-M^2}\;\f{1}{\Dbar{i}} \\[2mm]
\vcenter{\hbox{\includegraphics[width=1.5cm]{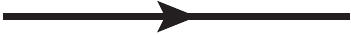}}} &=&
\vcenter{\hbox{\includegraphics[width=1.5cm]{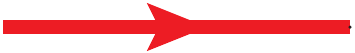}}} + 
\vcenter{\hbox{\includegraphics[width=2.5cm]{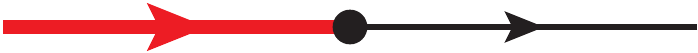}}}{},
\label{eq:dentadexp}
\eea
with
\bea
\label{eq:dentadexpA3}
\Delta_k(\denbar q_1)&=& \left(\denbar q_1^2-M^2\right)  -D_k(\denbar q_1) =
-2 \momq_1\cdot p_k-p_k^2+m_k^2-M^2\,,
\eea
where $M^2$ is an auxiliary squared mass scale. We have introduced a graphical representation, 
in which thick red lines depict the pure tadpole
propagators $(\momq_1^2-M^2)^{-2}$ and black ones to the original propagators $\Dbar{i}$.
This can be used recursively up to a fixed order $X+1$ in the tadpole propagators, 
in order to isolate the UV divergences and rational terms in pure tadpole integrals 
with one auxiliary scale $M^2$.

For a more compact notation, we introduce 
two operators $\bfS^{(1)}_X$ and $\bfF^{(1)}_X$, which generate, respectively, the pure tadpole 
terms up to order $X$ in a naive $1/q$ power counting of the integrand, and the higher-order remnant containing all original
propagators,
\footnote{The label $(i)$ in $\bfS^{(i)}_X$ and 
$\bfF^{(i)}_X$ refers to the chain of propagator denominators with loop momentum $\momq_i$,
on which these operators act exclusively, since in multi-loop diagrams there are several 
chains with different $\momq_i$.} 
\bea
\f{1}{\Dbar{k}(\momq_1)}&=& \lb \bfS_X^{(1)} + \bfF_X^{(1)} \rb
\f{1}{\Dbar{k}(\momq_1)}\,, \label{eq:dentadexpB}
\eea
with 
\bea
\label{eq:dentadexpC}
\bfS^{(1)}_X \frac{1}{D_k(\denbar q_1)}&=& 
\sum_{\sigma=0}^X
\frac
{\left[\Delta_k(\denbar q_1)\right]^{\sigma}}
{\left(\denbar q_1^2-M^2\right)^{\sigma+1}}\,,
\qquad
\bfF^{(1)}_X \frac{1}{D_k(\denbar q_1)}
\,=\,
\frac
{\left[\Delta_k(\denbar q_1)\right]^{X+1}}
{\left(\denbar q_i^2-M^2\right)^{X+1}}
\frac{1}{D_k(\denbar q_1)}\,{}.
\eea
 If $X$ is chosen as the superficial degree of divergence
and \eqref{eq:dentadexpB} applied to every propagator denominator along the loop,
all terms containing an original $\Dbar{k}$ are free from UV divergences.

For a full chain of propagators these operators are defined to give
\bea
\label{eq:chaintadexpA2}
\bfS^{(1)}_X 
\frac{1}{D_0(\denbar q_1)\cdots D_{N-1}(\denbar q_1)}&=& 
\sum_{\sigma=0}^{X}
\frac
{\Delta^{(\sigma)}(\denbar q_1)}
{\left(\denbar q_1^2-M^2\right)^{N+\sigma}}\,,\qquad \bfF^{(1)}_X=  1 - \bfS^{(1)}_X{},
\eea
where
\bea
\label{eq:chaintadexpB}
{\Delta^{(\sigma)}(\denbar q_1)}
&=& 
\sum_{\sigma_0=0}^\sigma
\ldots
\sum_{\sigma_0={N-1}}^\sigma
\prod_{k=0}^{N-1}
\left[\Delta_k(\denbar q_1)\right]^{\sigma_k}
\Bigg|_{\sigma_0+\dots+ \sigma_{N-1} = \sigma}\,
\eea
is a polynomial in the external momenta, the masses and the loop momentum.
Since the finite terms collected by $\bfF^{(1)}_X$ cancel in the difference \eqref{eq:RA1f},
we can compute the rational term of type $R_2$ as
\be
\ratamp{1}{\gamma}{}{}= \bfS^{(1)}_X (\ampbar{1}{\gamma}{}{}-\amp{1}{\gamma}{}{})
=\sum\limits_{\sigma=N}^{N+X}
\int\!\rd\momq_1\, 
\f{ \left(\barN(\momq_1) - \calN(q_1) \right) 
\,\Delta^{(\sigma)}(\momq_1)}{(\momq_1^2-M^2)^{\sigma} }{}. \label{eq:R21loop_comp}
\ee
The numerator in \eqref{eq:rtoneloopA} can be written in terms of loop momentum tensors up to rank $R$,
\bea
\label{eq:oneloopstructD3}
\bar \calN(\bar q_1)
&=&
\sum_{r=0}^R  \bar\calN_{\bar\mu_1\cdots \bar\mu_r}\,
\bar q_1^{\bar\mu_1}\cdots \bar q_1^{\bar\mu_r}\,.
\eea

\begin{figure}[t]
\newcommand{\diawidthoneloop}{1.9cm}
\newcommand{\ddimnumIL}{\!\!\!\!\!\!\!\!\!\!\!\!\!\!\!\!\!\!\!\!\!\!\!\!\!\!\!\!\! \text{\scriptsize $\numdim=\dendim$}\quad}
\newcommand{\fdimnumIL}{\!\!\!\!\!\!\!\!\!\!\!\!\!\!\!\!\!\!\!\!\!\!\!\!\!\!\!\!\! \text{\scriptsize $\numdim=4$}\quad}
\beas
\ratamp{1}{\gamma}{}{} \!\!&=&\!\!
\vcenter{\hbox{\includegraphics[width=\diawidthoneloop]{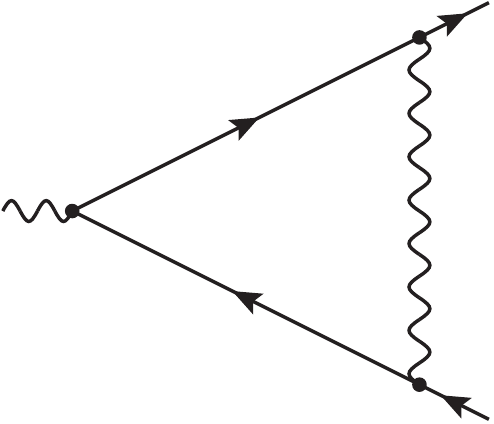}}}\ddimnumIL
-\vcenter{\hbox{\includegraphics[width=\diawidthoneloop]{QEDvtxIL}}}\fdimnumIL
\;=\;\vcenter{\hbox{\includegraphics[width=\diawidthoneloop]{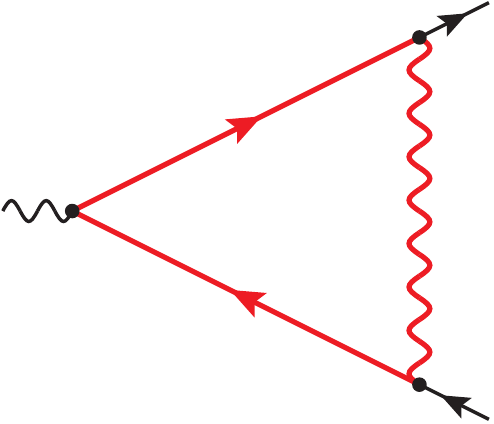}}}\ddimnumIL
-\vcenter{\hbox{\includegraphics[width=\diawidthoneloop]{QEDvtxILM}}}\fdimnumIL +
\underbrace{\vcenter{\hbox{\includegraphics[width=\diawidthoneloop]{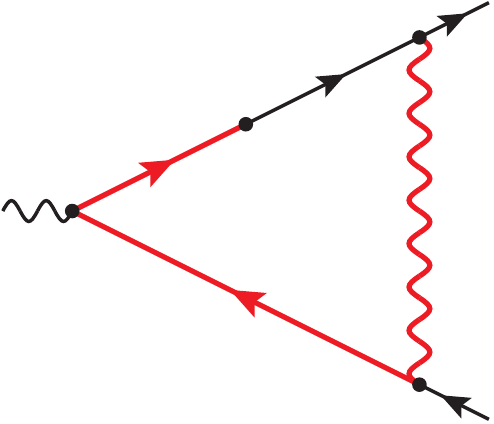}}}\ddimnumIL-
\vcenter{\hbox{\includegraphics[width=\diawidthoneloop]{QEDvtxILMI}}}\fdimnumIL}_{=\mathcal{O}(\eps)} \\ \!\!&+&\!\!
\ldots + \underbrace{\vcenter{\hbox{\includegraphics[width=\diawidthoneloop]{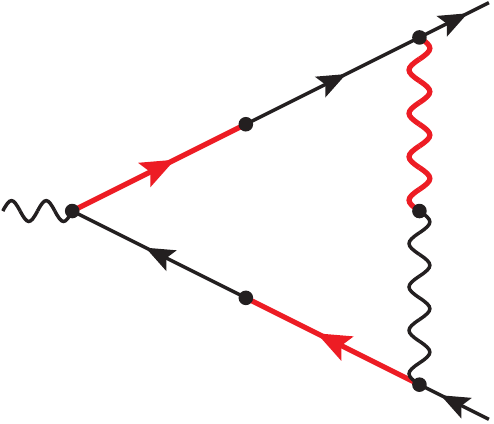}}}\ddimnumIL-
\vcenter{\hbox{\includegraphics[width=\diawidthoneloop]{QEDvtxILMVII}}}\fdimnumIL}_{=\mathcal{O}(\eps)}
\,=\, \bfS_0^{(1)} (\barA_{1,\gamma}-\calA_{1,\gamma}) + \mathcal{O}(\eps){}\\[-8mm]
\eeas
\caption{\label{fig:oneloop_example} $\tilN$-contribution for the one-loop vertex correction in QED. $\numdim$ is the numerator dimension.}
\end{figure}

Decomposing the extended numerators and including the 
resulting term \eqref{eq:R21loop_comp} into the loop numerator
yields
\bea
\!\!\!\!\!\!\!\!\!\!\!\!\!\!\!\!\!
\barN(\momq_1)\,\Delta^{(\sigma)}(\momq_1) &=& \sum\limits_{r=0}^{R+\sigma}
\barN^{(\sigma)}_{\bar\mu_1\cdots\bar\mu_r} \momq_1^{\bar\mu_1}\cdots\momq_1^{\bar\mu_r}, \qquad 
\calN(q_1)\,\Delta^{(\sigma)}(\momq_1) = \sum\limits_{r=0}^{R+\sigma}
\calN^{(\sigma)}_{\mu_1\cdots\mu_r} 
q_1^{\mu_1}\cdots q_1^{\mu_r} {}.\;\;
\eea
Hence, we can compute the $\tilN$-contributions from coefficients constructed in $\dendim$ and
four dimensions and fully $\dendim$-dimensional tensor integrals,
\be
\ratamp{1}{\gamma}{}{} =
\sum\limits_{\sigma=0}^{X}
\sum\limits_{r=0}^{R+\sigma}
 \left(\barN^{(\sigma)}_{\bar\mu_1\cdots\bar\mu_r} - \calN^{(\sigma)}_{\mu_1\cdots\mu_r} \right) 
\int\!\rd\momq\, 
\f{\momq^{\bar\mu_1}\cdots\momq^{\bar\mu_r}}
{(\momq^2-M^2)^{n} }{}, \label{eq:masterformulaR2_1loop} 
\ee
using $\momq\cdot{}p_j=q\cdot p_j$ for external four-momenta $p_j$ and
$A_{\mu_i} B^{\bar\mu_i}=A_{\mu_i} B^{\mu_i}$ for any vectors $A,B$.
Since the denominator decomposition \eqref{eq:dentadexpB} is exact and $M^2$-independent, and 
\mbox{$
\bfF^{(1)}_X (\ampbar{1}{\gamma}{}{}-\amp{1}{\gamma}{}{})=\mathcal{O}(\eps),
$}
the finite part of $\bfS^{(1)}_X (\ampbar{1}{\gamma}{}{}-\amp{1}{\gamma}{}{})$ and hence
$\ratamp{1}{\gamma}{}{}$ does also not depend on $M^2$.
The dependence on external momenta and masses resides exclusively in the $\barN^{(\sigma)}_{\bar\mu_1\cdots\bar\mu_r}$ and 
 $\calN^{(\sigma)}_{\mu_1\cdots\mu_r}$ in polynomial form, which proves that $\ratamp{1}{\gamma}{}{}$ is indeed a rational term.
We illustrate the cancellation of terms with original propagators and the resulting pure 
tadpole contribution for the one-loop vertex correction in QED in Fig.~\ref{fig:oneloop_example}. 

\subsection{One-loop sub-diagrams with numerator dimension $\numdim=4$}
In the case of a one-loop sub-diagram with loop momentum $\momq_1$ to a two-loop diagram, we encounter a subtlety, namely the fact 
that one external momentum flowing through the sub-diagram
is the second loop momentum $\momq_2$, which is $\dendim$-dimensional in the construction of the numerator $\barN(\momq_1,\momq_2)$ and 
four-dimensional in the construction of $\calN(q_1,q_2)$, but always $\dendim$-dimensional in the denominators.
This requires an extension of \eqref{eq:RA1f} in order to subtract all divergences arising in $\amp{1}{\gamma}{}{}(q_2)$.

In this section we therefore consider the renormalisation of $\ampbar{1}{\gamma}{}{}(\momq_2)$ and 
its variant $\amp{1}{\gamma}{}{}(q_2)$, where the argument refers to the momentum in the numerator of the integrand.
The denominators are in both cases constructed with $\momq_2$ in $\dendim$ dimensions.

In the case of a fully $\dendim$-dimensional one-loop amplitude the UV counterterm in the 
$\overline{\text{MS}}$-scheme, introduced in \eqref{eq:RA1d}, is derived as\footnote{We suppress the indices $\alpha_1,\alpha_2$ connecting the sub-diagram to the rest of the two-loop diagram,
$\ampbar{1}{\gamma}{}{}=\ampbar{1}{\gamma}{\bar\alpha_1\bar\alpha_2}{}$, etc.}
\be
\deltaZ{1}{\gamma}{}{}(\momq_2)=-\textbf{K}\, \ampbar{1}{\gamma}{}{}(\momq_2),
\ee
where the operator $\textbf{K}$ performs a Laurent expansion in $\eps$ and discards all terms of order $\eps^0$ and higher,
isolating the pole part of an amplitude. We compute the UV counterterm for the same amplitude with a four-dimensional 
numerator as
\bea
\textbf{K}\, \amp{1}{\gamma}{}{}(q_2) &=& \textbf{K}\, \bfS_X^{(1)} \amp{1}{\gamma}{}{}(q_2) = \textbf{K}\,\sum\limits_{r=0}^{R} 
\calN_{\mu_1\ldots\mu_r}(q_2)
\sum\limits_{\sigma=0}^{X}\int\!\rd\momq_1\, 
\f{\momq_1^{\bar\mu_1}\cdots\momq_1^{\bar\mu_r}\Delta^{(\sigma)}(\momq_1,\momq_2)}{(\momq_1^2-M^2)^{N+\sigma}} \\
&=& - \deltaZ{1}{\gamma}{}{}(q_2)  - \deltaZtilde{1}{\gamma}{}{}(\tilq_2){}, 
\eea
where the first term on the rhs is the 
usual $\overline{\text{MS}}$ counterterm
with $\momq_2 \to q_2$, and
the $\dendim$-dimensional $\momq_2=q_2+\tilq_2$ in the extended numerator leads to an additional 
counterterm
$\deltaZtilde{1}{\gamma}{}{}(\tilq_2) \propto \tilq_2^2/\eps$ stemming from scalar products $\tilq_1\tilq_2$ and 
$\tilq_2^2$ in $\Delta^{(\sigma)}$ for $X\geq 2$.
This is a direct consequence of the fact that
scalar denominators are still $\dendim$-dimensional. Hence the renormalised amplitude, i.e. one with all poles in $\eps$ subtracted, in 
$\numdim=4$ can be defined as
\be
\bfR \amp{1}{\gamma}{}{}(q_2) := (1-\bfK) \amp{1}{\gamma}{}{}(q_2) = \amp{1}{\gamma}{}{}(q_2)+ \deltaZ{1}{\gamma}{}{}(q_2) + \deltaZtilde{1}{\gamma}{}{}(\tilq_2)
\ee
and the renormalised $\dendim$-dimensional amplitude can be written as 
\be
\bfR \ampbar{1}{\gamma}{}{}(\momq_2) = \bfR \amp{1}{\gamma}{}{}(q_2) +
\ratamp{1}{\gamma}{}{}(q_2) +\mathcal{O}(\eps,\tilq) \label{eq:oneloopsubdiaDdf}
\ee
with the well-known one-loop rational term of type $R_2$ \cite{R2paper}.
{
\newcommand{\ddimid}{\!\!\!\!\!\!\!\!\!\!\!\!\!\!\!\!\!\!\!\!\!\!\!\!\!\!\!\!\! \;\;\,\quad \text{\fontsize{8}{8} $\numdim=d$}\qquad}
\newcommand{\fdimid}{\!\!\!\!\!\!\!\!\!\!\!\!\!\!\!\!\!\!\!\!\!\!\!\!\!\!\!\!\! \;\;\,\quad \text{\fontsize{8}{8} $\numdim=4$}\qquad}
\newcommand{\lorentzMUNUdd}{^{\!\!\!\!\!\!\!\!\!\!\! \bar\alpha_1}_{\!\!\!\!\!\!\!\!\!\!\! \bar\alpha_2}}
\newcommand{\lorentzMUNUff}{^{\!\!\!\!\!\!\!\!\!\!\! \alpha_1}_{\!\!\!\!\!\!\!\!\!\!\! \alpha_2}}
\newcommand{\lorentzMUNUddct}{^{ \bar\alpha_1}_{ \bar\alpha_2}}
\newcommand{\lorentzMUNUffct}{^{ \alpha_1}_{ \alpha_2}}

\begin{figure}[t]
\begin{center}
\bea
\bfR\;\left[ 
\vcenter{\hbox{\includegraphics[height=\diaheight]{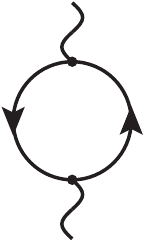}}}
\lorentzMUNUdd  
\right]_{\numdim\,=\,\dendim}
\hspace{-2mm}
&=& 
\left[ 
\vcenter{\hbox{\includegraphics[height=\diaheight]{QEDgpropIL}}}
\lorentzMUNUdd 
\;\;+\;\;
\vcenter{\hbox{\includegraphics[height=\diaheight]{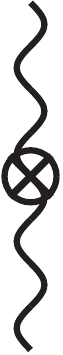}}}
\lorentzMUNUddct
\;\deltaZ{1}{\gamma}{}{}(\momq_2)\,
\right]_{\numdim\,=\,\dendim}
\hspace{-3mm}
\nonumber\\[4mm]
&=& 
\left[ \vcenter{\hbox{\includegraphics[height=\diaheight]{QEDgpropIL}}}
\lorentzMUNUff
\;\;+\;\;
\vcenter{\hbox{\includegraphics[height=\diaheight]{QEDgpropCT}}}
\lorentzMUNUffct
\;\Big(\deltaZ{1}{\gamma}{}{}(q_2)
+
\deltaZtilde{1}{\gamma}{}{} (\tilq_2)
+
\ratamp{1}{\gamma}{}{}(q_2)
\Big)
\right]_{\numdim\,=\,4}   
+\, \mathcal O(\eps)\;.
\nonumber
\eea
\end{center}
\caption{Graphical representation for a renormalised sub-diagram in numerator dimensions 
$\numdim=\dendim$ and
$\numdim=4$ for the case of a QED self-energy. The $\dendim$-dimensional external momentum $\momq_2$ is decomposed
into a four-dimensional part $q_2$ and a $(\dendim-4)$-dimensional one $\tilq_2=\momq_2-q_2$.
} 
\label{fig:rensubdiag}
\end{figure}
}
In renormalisable theories, $X\geq 2$ is only fulfilled for self-energies, in QED only for the photon self-energy. 
The renormalisation procedure for this case is depicted in Fig.~\ref{fig:rensubdiag}.
In all other cases $\deltaZtilde{1}{\gamma}{}{}(\tilq_2)=0$.

\section{Rational terms at two loops}
A generic irreducible two-loop diagram\footnote{Reducible diagrams can be computed by applying the previous 
discussion to each one-loop
sub-diagram $\gamma_1, \gamma_2$ separately, since
\bes
\bfR \ampbar{2}{\Gamma}{}{} = \bfR \ampbar{1}{\gamma_1}{}{} \,\cdot\, \bfR \ampbar{1}{\gamma_2}{}{}  {}.
\ees} 
$\Gamma$ can be decomposed into three chains, each consisting of a numerator $\barN^{(i)}(\momq_i)$ and a denominator
\bea
\label{eq:twoloopnotB}
\calD{i}&=&
D^{(i)}_0(\denbar q_i)\cdots
D^{(i)}_{N_i-1}(\denbar q_i)\,,
\qquad\mbox{with}\quad
D^{(i)}_a(\denbar q_i) \,=\, \left(\denbar q_i-p_{ia}\right)^2-m_{ia}^2\,,
\eea
as well as a function $\Gamma^{\bar\alpha_1\bar\alpha_2\bar\alpha_3}$ derived 
from the two connecting vertices $V_0$ and $V_1$,\footnote{The multi-indices
$\bar\alpha_i$ consist of two Lorentz or spinor indices.}
 \bea
\ampbar{2}{\Gamma}{}{}  &=& \vcenter{\hbox{\scalebox{1.}{\includegraphics[height=3cm]{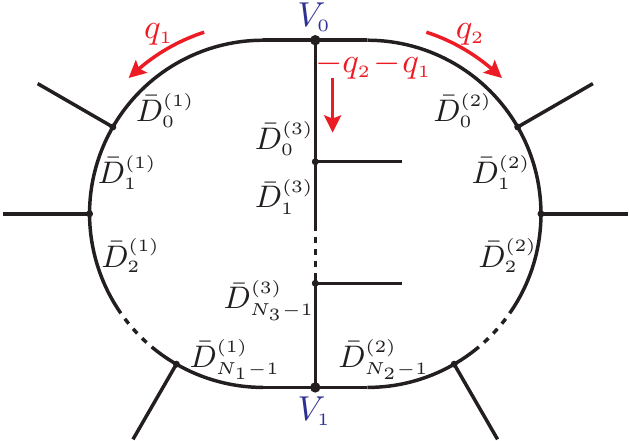}}}}{} =
\int\rd\momq_1\!\!\!
\int \rd\momq_2\!
\frac{
\bar\calN(\momq_1,\momq_2)}
{\calD{1}\,\calD{2}\,\calD{3}}
\nonumber \\[2mm]
&=&
\int\rd\momq_1\!\!\!
\int \rd\momq_2\!
\frac{
\bar\Gamma^{\bar\alpha_1\bar\alpha_2\bar\alpha_3}(\momq_1,\momq_2)\,
\bar\calN^{(1)}_{\bar\alpha_1}(\momq_1)\,
\bar\calN^{(2)}_{\bar\alpha_2}(\momq_2)\,
\bar\calN^{(3)}_{\bar\alpha_3}(\momq_3)}
{\calD{1}\,\calD{2}\,\calD{3}} \Bigg|_{\momq_3=-\momq_2-\momq_1} \label{eq:twoloopnotA1}
\eea 
There are three sub-diagrams $\gamma_i$, each constructed from two chains $j,k$ and the connecting function 
$\Gamma^{\bar\alpha_1\bar\alpha_2\bar\alpha_3}$, where $i|jk$ is a partition of $123$. The amplitude of a sub-diagram $\gamma_i$
is given by
\bea
\label{eq:subdiagnotA}
\ampbar{1}{\gamma_i}{\bar\alpha_i}{\bar q_i} 
&=&
\int\rd\momq_j\,
\frac{
\bar\Gamma^{\bar\alpha_1\bar\alpha_2\bar\alpha_3}\,
\bar\calN^{(j)}_{\bar\alpha_j}(\momq_j)\,
\bar\calN^{(k)}_{\bar\alpha_k}(\momq_k)}
{\calD{j}\,\calD{k}}\Bigg|_{\momq_k=-\momq_i-\momq_j}\,,
\eea
and the insertion into its complement $\Gamma/\gamma_i$ can be written as
\bea
\label{eq:subdiagnotB}
\ampbar{2}{\Gamma}{}{}&=&
\ampbar{1}{\gamma_i}{}{}\cdot
\ampbar{1}{\Gamma/\gamma_i}{}{}
\,=\,
\int\rd\momq_i\,
\ampbar{1}{\gamma_i}{\bar\alpha_i}{\momq_i}\,
\frac{\bar\calN^{(i)}_{\bar\alpha_i}(\momq_i)}{\calD{i}}
\,{}.
\eea
Formula \eqref{eq:masterformula}, which constructs a renormalised $\dendim$-dimensional amplitude from ingredients with four-dimensional numerators and insertions of UV counterterms and known one-loop rational terms,
implicitly defines the yet unknown remnant $\ratamp{2}{\Gamma}{}{}$ as
\bea
\label{eq:deltaAtwo}
\ratamp{2}{\Gamma}{}{}
&=&\lb\ampbar{2}{\Gamma}{}{} + \sum  \limits_{\gamma}  \deltaZ{1}{\gamma}{}{} \cdot \ampbar{1}{\Gamma/\gamma}{}{}\rb
-\lb \amp{2}{\Gamma}{}{} + 
\sum  \limits_{\gamma} \lb \deltaZ{1}{\gamma}{}{}+\deltaZtilde{1}{\gamma}{}{} + \ratamp{1}{\gamma}{}{} \rb \cdot \amp{1}{\Gamma/\gamma}{}{} \rb {}.
\eea
We will show in the following that $\ratamp{2}{\Gamma}{}{}$
is polynomial in all external momenta and masses, which makes it a rational term that can be written as a local counterterm.

We distinguish two cases, diagrams with superficial degree of divergence $X(\Gamma)<0$ (no global divergence) and diagrams
with $X(\Gamma)>0$ (global divergence). 

\subsection{Diagrams without global divergence} \label{sec:twoloop-wo_global}

Two-loop diagrams with $X(\Gamma)<0$ can have at most one divergent sub-diagram $\gamma_i$ (see e.g.~\cite{R2paper}), i.e.~a sub-diagram with $X(\gamma_i)\geq 0$.
The renormalised amplitude in $\numdim=d$ dimensions is given as
\bea
\label{eq:noglobdivA}
\bfR \ampbar{2}{\Gamma}{}{} &=&
\ampbar{2}{\Gamma}{}{} -
\left(\bfK\ampbar{1}{\gamma_i}{}{}\right)\cdot \ampbar{1}{\Gamma/\gamma_i}{}{}
\,=\,
\left(\ampbar{1}{\gamma_i}{}{}+\deltaZ{1}{\gamma_i}{}{} \right)
\cdot \ampbar{1}{\Gamma/\gamma_i}{}{} .
\eea
Since the sub-diagram $\gamma_i$ plus its UV counterterm,
and also its counterpart $\Gamma/\gamma_i$ are 
free from any divergence
we can discard any $\ord(\eps)$ contributions both in 
the renormalised sub-diagram $\gamma_i$ and $\Gamma/\gamma_i$. 
Hence, we construct the numerator of $\Gamma/\gamma_i$ in $\numdim=4$,
leading to\footnote{In this step, we also project the Lorentz indices connecting $\gamma_i$ and $\Gamma/\gamma_i$ to four dimensions, $\bar\alpha_i\to\alpha_i$.}
\bea
\label{eq:noglobdivB1}
\bfR\ampbar{2}{\Gamma}{}{} &=&
\left(\ampbar{1}{\gamma_i}{}{}+\deltaZ{1}{\gamma_i}{}{}\right)
\cdot \amp{1}{\Gamma/\gamma_i}{}{} +\ord(\eps){}.
\eea
The UV-subtracted $\gamma_i$ can be expressed with its amplitude constructed in $\numdim=4$
using \refeq{eq:oneloopsubdiaDdf},
\bea
\label{eq:noglobdivB2}
\bfR\ampbar{2}{\Gamma}{}{}
&=&
\left(\amp{1}{\gamma_i}{}{}
+\deltaZ{1}{\gamma_i}{}{} 
+\deltaZtilde{1}{\gamma_i}{}{}
+\ratamp{1}{\gamma_i}{}{} \right) 
\cdot \amp{1}{\Gamma/\gamma_i}{}{}+\ord(\eps),
\eea
which means that for $X(\Gamma)<0$ we find
\bea
\label{eq:noglobdivD}
\ratamp{2}{\Gamma}{}{} &=& 0\,.
\eea
This shows that non-vanishing two-loop rational terms of UV origin occur only in 
the finite set of diagrams with a global divergence.

\subsection{Diagrams with global divergence}
In this section we show that $\ratamp{2}{\Gamma}{}{}$ in \eqref{eq:masterformula} is also a rational term for $X(\Gamma)\geq 0$. 
Our starting point is
the difference \eqref{eq:deltaAtwo}, which gives the remaining $\tilN$-contribution after
the subtraction of the sub-divergences from $\ampbar{2}{\Gamma}{}{}$ and 
$\amp{2}{\Gamma}{}{}$, and the restoration of the $\tilN$-contributions stemming from 
the one-loop sub-diagrams of $\Gamma$.
We show that \eqref{eq:deltaAtwo} can be fully computed from massive tadpoles with one auxiliary scale $M^2$, 
of which the result is independent. We then conclude, similar to the one-loop case, that \eqref{eq:deltaAtwo} is polynomial in the external momenta and
masses, and hence a rational term.

Our strategy is to apply a tadpole decomposition \eqref{eq:chaintadexpA2} to the three chains 
in $\ampbar{2}{\Gamma}{}{}$ and $\amp{2}{\Gamma}{}{}$
and their subtraction terms to high enough powers in $1/q_i$, such that all terms captured by the operators $\bfF_{X_i}^{(i)}$, in particular all terms involving original propagators, 
are free from global divergences
and hence cancel exactly
in \eqref{eq:deltaAtwo}
according to \refeq{eq:noglobdivD}.

\begin{figure}[t]
\begin{center}
\bea
&&\bfR\,
\left[\;
\vcenter{\hbox{\includegraphics[width=\diawidth]{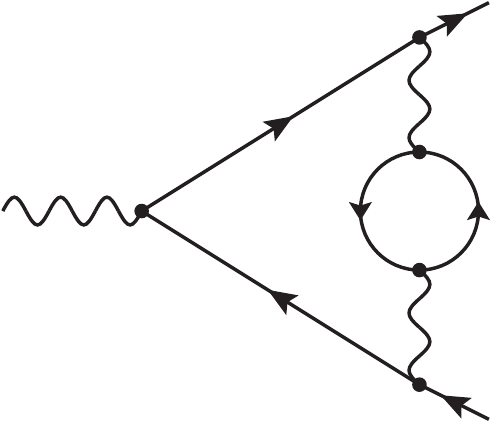}}}\;\;\right]_{\numdim\,=\,\dendim}
\hspace{-3mm}
= \!\! 
\left[\;
\vcenter{\hbox{\includegraphics[width=\diawidth]{QEDvtxIILoop}}}  
\;\;+\;
\vcenter{\hbox{\includegraphics[width=\diawidth]{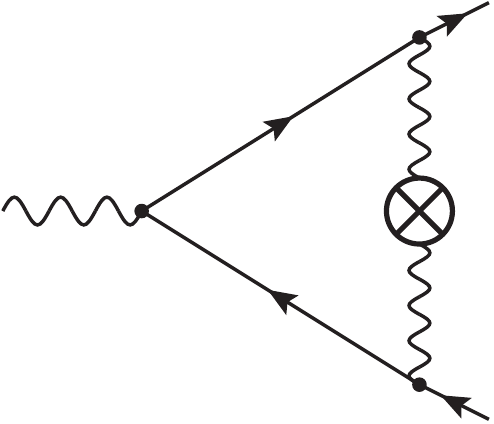}}}{
\deltaZ{1}{\gamma_i}{}{} } 
\;+\;
\vcenter{\hbox{\includegraphics[width=\diawidth]{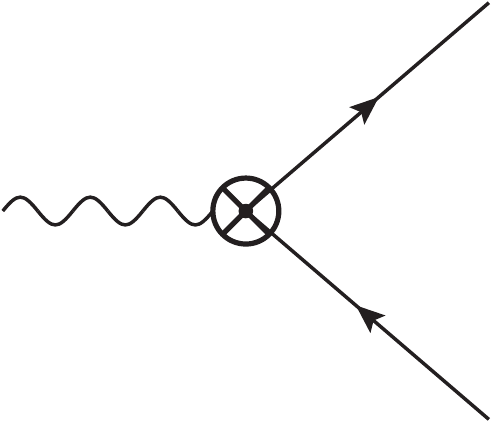}}} \hspace{-5mm}\deltaZ{2}{\Gamma}{}{} 
\;\right]_{\numdim\,=\,\dendim}
\nonumber\\
&&=\,
\left[\;
\vcenter{\hbox{\includegraphics[width=\diawidth]{QEDvtxIILoop}}}  
\;\;+\;
\vcenter{\hbox{\includegraphics[width=\diawidth]{QEDvtxILoopCT}}}{
\left(\deltaZ{1}{\gamma_i}{}{}+ \deltaZtilde{1}{\gamma_i}{}{} 
+\ratamp{1}{\gamma_i}{}{}\right)
} 
\;+\;
\vcenter{\hbox{\includegraphics[width=\diawidth]{QEDvtxCT}}} \hspace{-5mm}
\left(\deltaZ{2}{\Gamma}{}{} 
+\ratamp{2}{\Gamma}{}{}\right)
\;\right]_{\numdim\,=\,4}
\nonumber
\eea
\end{center}
\caption{Graphical representation of the renormalisation formula (first line) in $\dendim$ dimensions
for a two-loop QED diagram with a single sub-divergence, and the master formula (second line) to compute it from 
four-dimensional amplitudes with universal counterterm insertions.} 
\label{fig:twolooprat}
\end{figure}

To this end we express the rhs of \eqref{eq:deltaAtwo} with the linear operators $\bfK$ and $\tilde\bfK$, 
\be
\label{eq:deltaAtwoOp}
\lb\ampbar{2}{\Gamma}{}{} + \sum  \limits_{i}  \lb-\bfK\ampbar{1}{\gamma_i}{}{}\rb \cdot \ampbar{1}{\Gamma/\gamma_i}{}{}\rb
-\lb \amp{2}{\Gamma}{}{} + 
\sum  \limits_{i} \lb (-\bfK+ \tilde\bfK) \amp{1}{\gamma_i}{}{} \rb \cdot \amp{1}{\Gamma/\gamma}{}{} \rb
=: \bfKtildeloc\ampbar{2}{\Gamma}{}{},
\ee
where
\be
\tilde\bfK\amp{1}{\gamma_i}{}{} := \bfR \ampbar{1}{\gamma_i}{}{} - \bfR \amp{1}{\gamma_i}{}{}  
 \equiv
\ratamp{1}{\gamma_i}{}{}.
\ee
The operator $\bfKtildeloc$, which constructs $\ratamp{2}{\Gamma}{}{}$ is in turn linear.
The divergent parts of a two-loop 
diagram and its three sub-diagrams can be isolated by applying 
independent tadpole decompositions \eqref{eq:chaintadexpA2}
to the three chains. Each chain $i$ is expanded to the maximum of the global divergence of $\Gamma$ and the divergences of the
sub-diagrams $\gamma_j$, $\gamma_k$ of which it is a part,\footnote{Again $i | jk$ is a partition of 123.}
\bea
X_i (\Gamma) &=&
\text{max}\left\{ X(\Gamma), X(\gamma_j),
X(\gamma_k)\right\}\,.
\label{eq:powercountingDirectImp}
\eea
Applying this operation to \eqref{eq:deltaAtwoOp} gives
\bea
\label{eq:tadexptwoloopB}
\bfKtildeloc\ampbar{2}{\Gamma}{}{} 
&=& 
\bfKtildeloc
\left(\bfS^{(1)}_{X_1}+\bfF^{(1)}_{X_1}\right)
\left(\bfS^{(2)}_{X_2}+\bfF^{(2)}_{X_2}\right)
\left(\bfS^{(3)}_{X_3}+\bfF^{(3)}_{X_3}\right)
\ampbar{2}{\Gamma}{}{}\\
&=& 
\bfKtildeloc\ampbar{2}{\Gamma_\tad}{}{} 
+
\bfKtildeloc\ampbar{2}{\Gamma_\rem}{}{}\,, 
\eea
with a pure tadpole term and remnant terms
\bea
\label{eq:tadexptwoloopD}
\ampbar{2}{\Gamma_\tad}{}{} 
&=&
\bfS^{(1)}_{X_1}
\bfS^{(2)}_{X_2}
\bfS^{(3)}_{X_3}
\ampbar{2}{\Gamma}{}{}\,, 
\\
\label{eq:tadexptwoloopE}
\ampbar{2}{\Gamma_\rem}{}{} 
&=&
\bfF^{(1)}_{X_1}
\bfF^{(2)}_{X_2}
\bfF^{(3)}_{X_3}
\ampbar{2}{\Gamma}{}{} 
+
\sum_{i=1}^3
\bfS^{(i)}_{X_i}
\bfF^{(j)}_{X_j}
\bfF^{(k)}_{X_k}
\ampbar{2}{\Gamma}{}{} 
+
\sum_{i=1}^3 
\bfF^{(i)}_{X_i}
\bfS^{(j)}_{X_j}
\bfS^{(k)}_{X_k}
\ampbar{2}{\Gamma}{}{} 
\,.
\eea
All terms that enter \refeq{eq:tadexptwoloopE}
involve at least one $\bfF$ operator, which reduces the superficial degree of divergence 
$X(\Gamma)$, as well as $X(\gamma_j)$ and
$X(\gamma_k)$,
by $X_i(\Gamma)+1$. Thus the remnant part \refeq{eq:tadexptwoloopE}
is free from global divergences, and the discussion in section~\ref{sec:twoloop-wo_global} implies\footnote{Note that the operators $\bfS$ and $\bfF$ either 
fully reconstruct UV counterterms and rational terms or give zero, e.g.
$\bfK\lb \bfS^{(j)}_{X_j}\bfS^{(k)}_{X_k}\ampbar{1}{\gamma_i}{}{}\rb=-\deltaZ{1}{\gamma_i}{}{}$ and
$\bfK\lb\bfS^{(j)}_{X_j}\bfF^{(k)}_{X_k}\ampbar{1}{\gamma_i}{}{}\rb=0$ and 
$\bfK\lb\bfF^{(j)}_{X_j}\bfF^{(k)}_{X_k}\ampbar{1}{\gamma_i}{}{}\rb=0$. 
}
\\[-8mm]
\bea
\label{eq:tadexptwoloopG}
\ratamp{2}{\Gamma_\rem}{}{} &=&
\bfKtildeloc\, \amp{2}{\Gamma_\rem}{}{}
= 0 {}.
\eea
Hence $\ratamp{2}{\Gamma}{}{}$ can be fully 
computed from tadpole integrals with one auxiliary scale $M^2$,
\bea
\ratamp{2}{\Gamma}{}{}=
\bfKtildeloc\, \amp{2}{\Gamma}{}{}&=&
\bfKtildeloc\, 
\prod\limits_{n=1}^{3}\bfS_{X_n}^{(n)}\amp{2}{\Gamma}{}{}
= \lb
\prod\limits_{n=1}^{3}\bfS_{X_n}^{(n)}
\ampbar{2}{\Gamma}{}{} 
+ \sum  \limits_{i}  \deltaZ{1}{\gamma_i}{}{} \cdot \bfS^{(i)}_{X_i}\ampbar{1}{\Gamma/\gamma_i}{}{}\rb  \nonumber\\
 & &
-\lb 
\prod\limits_{n=1}^{3}\bfS_{X_n}^{(n)}\amp{2}{\Gamma}{}{} + 
\sum  \limits_{i} \lb \deltaZ{1}{\gamma_i}{}{} + \deltaZtilde{1}{\gamma_i}{}{} 
+\ratamp{1}{\gamma_i}{}{}\rb \cdot \bfS^{(i)}_{X_i}\amp{1}{\Gamma/\gamma}{}{} \rb.\nonumber
\eea
Here we use that the operators $\bfS_{X_i}^{(i)}$ fully capture the UV counterterms and rational terms in the sub-diagrams due to the definition \eqref{eq:powercountingDirectImp} of the order of the tadpole decomposition, \ie
\be \bfK\lb \bfS^{(j)}_{X_j}\bfS^{(k)}_{X_k}\ampbar{1}{\gamma_i}{}{}\rb=-\deltaZ{1}{\gamma_i}{}{}, \qquad
\bfKtilde\lb \bfS^{(j)}_{X_j}\bfS^{(k)}_{X_k}\amp{1}{\gamma_i}{}{}\rb=\ratamp{1}{\gamma_i}{}{}. \ee
Because of the exactness of the tadpole decomposition \refeq{eq:tadexptwoloopB}
and \eqref{eq:tadexptwoloopG}, the result is independent of $M^2$
and polynomial in external momenta and internal masses, and hence indeed a rational term. 
A sample diagram and its renormalisation
in $\dendim$ dimensions as well as its computation via the master formula \eqref{eq:masterformula} are depicted in 
Fig.~\ref{fig:twolooprat}. 
\section{Results for QED}
In this section we present the two-loop rational terms in QED. They are calculated with QGRAF~\cite{QGRAF}, Q2E and EXP~\cite{Seidensticker:1999bb,Harlander:1997zb} 
and MATAD~\cite{MATAD}. We start from the Lagrangian
\be
\mathcal L_{QED} = \bar{\psi}(i\gamma^\mu D_\mu - m)\psi -\frac{1}{4}F_{\mu\nu}F^{\mu\nu} - \frac{1}{2\lambda} (\partial^\mu A_\mu)^2{},\qquad D_\mu = \partial_\mu - i e A_\mu
\ee
and set the fermion mass $m$ and the gauge parameter $\lambda=1$. The rational terms and the additional 
counterterm $\tilde{Z}$ for the photon self-energy are\\[3mm]
\begin{tabular}{lll}
$\vcenter{\hbox{\includegraphics[width=0.15\textwidth]{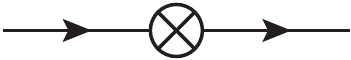}}}$
& $\ratamp{1}{e}{}{} =\frac{ie^2}{16\pi^2}  \left[ -1  \right]\slashed p$,
& $\ratamp{2}{e}{}{} =  \frac{ie^4}{(16\pi^2)^2}  \left[  {\frac{19}{18\eps}}        + \frac{247}{108}    \right] \slashed p$ \\[8mm]
$\vcenter{\hbox{\includegraphics[width=0.15\textwidth]{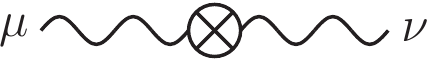}}}$
& $\ratamp{1}{\gamma}{\mu\nu}{} = \frac{ie^2}{16\pi^2} \left[ 
 \frac{2}{3} p^2\right] g^{\mu \nu} $,\!\!
& $\ratamp{2}{\gamma}{\mu\nu}{} = \frac{ie^4}{(16\pi^2)^2} \left[ P^{\mu\nu}
\left( {\frac{2}{3\eps}} - \frac{71}{18}  \right) 
 + g^{\mu \nu} p^2 \left( - \frac{11}{12} \right)  \right]$\\[8mm]
& $\deltaZtilde{1}{\gamma}{\mu\nu}{} = \frac{i e^2}{16\pi^2} \left[ { \frac{2}{3\eps}\tilde{p}^2}
\right] g^{\mu \nu} $ & \\ 
$\vcenter{\hbox{\includegraphics[width=0.15\textwidth]{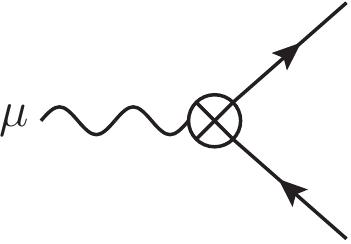}}}$
& $\ratamp{1}{e e \gamma}{}{} =   \frac{ie^3}{16\pi^2} \left[ -2  \right] \gamma^{\mu}   $,
& $\ratamp{2}{e e \gamma}{}{} =   \frac{ie^5}{(16\pi^2)^2}   \left[ 
{\frac{13}{9\eps}}           +  \frac{191}{27}  \right]\gamma^{\mu} $\\[8mm]
$\vcenter{\hbox{\includegraphics[width=0.15\textwidth]{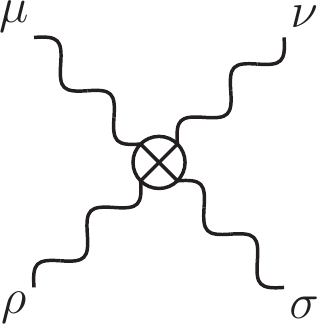}}}$
& $\ratamp{1}{4\gamma}{\mu\nu\rho\sigma}{} =  \frac{ie^4}{16\pi^2} \left[\frac{4}{3}\right] G^{\mu\nu\rho\sigma}
$, 
& $\ratamp{2}{4\gamma}{\mu\nu\rho\sigma}{} =  \frac{ie^6}{(16\pi^2)^2} \left[-3  \right] 
G^{\mu\nu\rho\sigma}$
\end{tabular}\\[5mm]
with $P^{\mu\nu}=p^{\mu}p^{\nu} - g^{\mu \nu} p^2$ and $G^{\mu\nu\rho\sigma}=g^{\mu \nu} g^{\rho \sigma} + g^{\mu \rho}g^{\nu \sigma} + g^{\mu \sigma} g^{\nu \rho}$. 
We find that the two-loop $\tilN$-contributions are indeed polynomial in the external momentum $p$ and hence are rational terms.
The full $m$ and $\lambda$-dependence of these terms is presented in \cite{R2paper}.
\section{Conclusions}
We have presented an extension of the renormalisation procedure for $\dendim$-dimensional amplitudes at two-loop level,
such that the numerators of all Feynman diagrams can be constructed in four dimensions. 
We showed that the contributions stemming from the interplay
of $(\dendim-4)$-dimensional numerator with UV divergences can be reconstructed through a finite set of local counterterms, and presented this set
for the case of QED. This constitutes an important building block for numerical two-loop calculations.

\section*{Acknowledgments}
We thank J.-N.~Lang for useful discussions. S.P., H.Z., M.Z. acknowledge support
from the Swiss National Science Foundation (SNF) under contract
BSCGI0-157722.
M.Z. acknowledges support by the Swiss National Science Foundation (Ambizione grant PZ00P2-179877).

\bibliographystyle{JHEP}

\providecommand{\href}[2]{#2}\begingroup\raggedright\endgroup

\end{document}